\documentclass[review]{elsarticle}

\usepackage{lineno,hyperref}
\usepackage{soul}
\usepackage[usenames, dvipsnames]{color}
\modulolinenumbers[5]

\journal{Chemical Physics Letters}

\begin{document}
\begin{frontmatter}
\title{Potential energy surfaces of the low-lying electronic states of the Li + LiCs system}

\author{P. Jasik$^{a,}$\fnref{email}}
\author{T. Kilich$^a$}
\author{J. Kozicki$^b$}
\author{J.E. Sienkiewicz$^{a,}$\fnref{email}}

\address[ftims]{Faculty of Applied Physics and Mathematics,
Gda\'nsk University of Technology, Gda\'nsk, Poland}

\address[wilis]{Faculty of Civil and Environmental Engineering,
Gda\'nsk University of Technology, Gda\'nsk, Poland}

\fntext[email]{Corresponding authors: patryk.jasik@pg.edu.pl (P.~Jasik), jes@mif.pg.gda.pl (J.E.~Sienkiewicz)}


%\cortext[corr_Jasik]{Corresponding authors}
%\ead{p.jasik@mif.pg.gda.pl}%, jes@mif.pg.gda.pl}


\begin{abstract}
{\it Ab initio} quantum chemistry calculations are performed for the mixed alkali triatomic system. Global minima of the ground and first excited doublet states of the trimer are found and Born-Oppenheimer potential energy surfaces of the Li atom interacting with the LiCs molecule were calculated for these states. The lithium atom is placed at various distances and bond angles from the lithium-caesium dimer. Three-body nonadditive forces of the Li$_2$Cs molecule in the global minimum are investigated. Dimer-atom interactions are found to be strongly attractive and may be important in the experiments, particularly involving cold alkali polar dimers.
\end{abstract}

\begin{keyword}
potential energy surfaces\sep Li2Cs\sep triatomic system\sep atom-molecule collisions\sep three-body nonadditive forces 
%\texttt{elsarticle.cls}\sep \LaTeX\sep Elsevier \sep template
\PACS 31.15.A-\sep 31.15.ae\sep 31.50.Bc\sep 31.50.Df\sep 33.15.Dj\sep 34.20.-b
%\MSC[2010] 00-01\sep  99-00
\end{keyword}

\end{frontmatter}

%\linenumbers

\section{Introduction}

There is a growing demand for data concerning heteronuclear alkali-metal dimers interacting with an alkali-metal atom. The investigations are carried out in the three main streams: the electronic structure properties, the dissociation or fragmentation processes and the association processes. The very extensive experimental and theoretical studies on the heteronuclear and homonuclear alkali trimers were given by the group of Ernst \cite{2008Nagl, 2008Nagl-1, 2008Hauser, 2011Hauser, 2011Giese}, where alkali quartet trimers formed on helium nanodroplets are probed by one-color femtosecond photoionization spectroscopy. The observation of predissociation in the mixed alkali trimer clusters was reported by the experimental group of W\"oste \cite{1998Vadja}. The same group presented the coherent control of alkali cluster fragmentation dynamics \cite{2003Lindinger}. The polar dimer gas can be controlled in optical traps and eventually converted into the quantum degenerate ultracold dipolar gas \cite{2015Moses}. Triatomic systems attract considerable attention in the context of experiments involving cold and ultracold molecules \cite{2012Ulmanis}. Efficient control of such three-body interactions requires detailed knowledge of low-lying interatomic potentials. Among other data, the potential energy surfaces (PESs) for a dimer interacting with an atom are of growing interest since they may help to describe elastic and inelastic atom-molecule collisions \cite{2010Soldan, 2010Zuchowski}, particularly when experiments on magnetic tuning of Feshbach resonances \cite{2008Jing, 2017Kato} and three-body recombination \cite{2013Harter,2016Moses} are considered.

In this paper, we study the interaction between the lithium atom and the lithium-caesium dimer. We also calculate the global minima of the ground ($1^2$A$'$) and first excited ($2^2$A$'$) doublet states of the Li$_2$Cs trimer, as well as three-body nonadditive contribution for the minimum of the ground state.

Until now, this system was very scarcely studied. To the authors' knowledge the first study concerning this system was performed in 1982 by Richtsmeier {\it et al.} \cite{1982Richtsmeier}, where energies of the optimized geometries for the linear LiCsLi in D$_{\infty h}$ and CsLiLi in C$_{\infty v}$ symmetries, as well as nonlinear Li$_2$Cs trimer in the C$_{2v}$ symmetry were presented. This study was carried out by means of the diatomics-in-molecules (DIM) approximation and used empirically evaluated integrals. In turn, \. Zuchowski and Hutson \cite{2010Zuchowski} modeled the reactions involving pairs of the alkali metal dimers. Using the multireference average-quadratic coupled-cluster method (AQCC), they found the global minima of mixed alkali-metal trimers, including Li$_2$Cs.

\section{Computational method}
In our calculations, Li + LiCs is considered as an effective three-electron system. Each n-electron atom is replaced by one valence electron and the effective core consisting of a nuclei and n-1 electrons. Since our theoretical approach has been already presented in a few earlier papers (e.g. \cite{2006Jasik, 2007Jasik, 2013Miadowicz, 2013Lobacz, 2015Wiatr}), here we give only salient details concerning pseudopotentials and atomic basis sets which differ from those used in our earlier calculations involving lithium and caesium atoms \cite{2006Jasik, 2013Szczepkowski}. The calculations are based on the multireference singles and doubles configuration interaction with Davidson correction (MRCISD+Q) method with atomic effective core potentials and core-polarization potentials, which enables us to treat only three valence electrons explicitly. The full configuration interaction (FCI) method is used to account for the correlation missing in MRCI calculations. An augmented atomic orbital basis allows to obtain reliable Born-Oppenheimer (BO) adiabatic potentials of several molecular states.

All calculations of the BO adiabatic potential energy curves (PECs) and surfaces are performed by means of the MOLPRO program package \cite{MOLPRO1, MOLPRO2}.

The core electrons of the caesium atom are represented by the ECP54SDF pseudopotential \citep{1982VonSzentpaly} and core-polarization potential with the values of dipole polarizability and cut-off parameter taken as $\alpha_D = 15.1~a_0^3$ and $\rho = 0.17~a_0^{-2}$, respectively. In the case of s and p functions, we use the basis set for caesium which comes with the ECP54SDF pseudopotential. For the d and f functions, we use the def2-QZVPPD basis set \cite{2010Rappoport}. Additionally, these basis sets are augmented by the six s functions with the given exponential coefficients of the Gaussian Type Orbitals (GTO) (2.055983, 1.188777, 0.687355, 0.009778, 0.005059, 0.002617), the four p functions (0.695867, 0.293629, 0.004186, 0.001830) and the two d functions (11.281944, 0.003919). In turn, for Li, the core electrons are represented by the ECP2SDF pseudopotential \cite{1982Fuentealba} with its respective core-polarization potential parameters ($\alpha_D = 0.1915~a_0^3$ and $\rho = 0.831~a_0^{-2}$). Basis set constructed for this pseudopotential is augmented by the six s functions (392.169555, 77.676373, 15.385230, 0.010159, 0.003894, 0.001493) and the four p functions (19.845562, 4.076012, 0.007058, 0.002598). Additionally, for d and f functions we use the cc-pV5Z basis set \cite{2011Prascher} augmented by the two d functions (1.043103, 0.026579). We check the quality of our basis sets by performing the CI calculations for the ground states of lithium and caesium atoms as well as for the several excited states of both atoms. The potential energy surfaces for the interaction between the LiCs dimer and the Li atom were computed using the multiconfigurational self-consistent field/complete active space self-consistent field (MCSCF/CASSCF) method to generate the orbitals for the subsequent configuration interaction calculations. The corresponding active space involves the molecular orbitals build from the 6s and 6p valence orbitals of caesium as well as 2s and 2p valence orbitals of lithium. Altogether the active space consists of 5 states in A$'$ and 2 states in A$''$ irreducible representations.

In order to determine the quality of our choice, we run dimer calculations for the aforementioned basis sets and pseudpotentials. Our calculated values of the LiCs dimers ground state dissociation energy (D$_e$ = 6090.259 cm$^{-1}$) and bond length (R$_e$ = 3.6160 \AA) are in good agreement with experimental \citep{2007Staanum} values (D$_e$ = 5875.455 cm$^{-1}$ and R$_e$ = 3.6681 \AA).

In order to test the active space and amount of correlation energy that is missing in MRCISD+Q computations, we run FCI calculations, both for the atomization energy of the Li$_2$Cs trimer and for minima. No difference between the MRCI and FCI results in the atomic limit was found, while the non-zero FCI corrections to atomization and dissociation energies are respectively specified in Tables 1 and 2.

Following Soldan {\it et al.} \cite{2003Soldan}, we decompose the three-atom interaction potential at the minimum of the ground state into a sum of additive (V$_{dimer}$) and nonadditive (V$_3$) contributions. Such decomposition can be expressed as

\begin{equation}\label{eq:tbnaf}
V_{trimer}(r_{12},r_{13},r_{23}) = \sum_{i<j} V_{dimer}(r_{ij}) + V_3(r_{12},r_{13},r_{23}).
\end{equation}

To describe the potential energy surfaces of Li + LiCs system we introduce two geometries written in the Z-matrix coordinates. The first geometry starts from the linear LiCs-Li system, where potential energy surfaces are calculated as a function of two variables. The first of these variables is the distance R between the Li atom and the Cs atom of the LiCs dimer, while the second variable is the bond angle $\theta$ as is shown in Fig \ref{LCL}. Here, the Li atom approaches the Cs atom of the dimer. The second geometry, as shown in the geometric scheme in Fig. \ref{CLL}, is analogical and starts from the linear Li-LiCs system ($\theta = 180^o$), where the Li atom approaches the dimer from the side on which the other Li atom is situated. In both geometries, the bond length R$_e$ = 3.6681 \AA \, was chosen as it is the experimental equilibrium distance of the LiCs dimer in the ground state \cite{2007Staanum}. Calculations of the BO potential energy surfaces are performed for R in the range from 4 a$_0$ to 100 a$_0$ and from 3.4 a$_0$ to 100 a$_0$ for geometries LiCs-Li and Li-LiCs, respectively, with the various steps adjusted to the atom-molecule distance. The bond angles are changed from 5$^o$ to 180$^o$ with the step equal to 5$^o$.

\section{Results and discussion}
Using the procedure described in the previous chapter we found the global minima of the ground and first excited doublet A$'$ states of the Li$_2$Cs trimer. We also calculated the FCI correction to atomization energies. Our results, compared with those of \. Zuchowski and Hutson \citep{2010Zuchowski} are shown in Table \ref{global}. Within the accuracy of one significant digit (that is given in \citep{2010Zuchowski}), our geometry of the ground state agrees perfectly with the recent study. The result of Richtsmeier {\it et al.} \citep{1982Richtsmeier} seems to overestimate the Cs-Li bond length and underestimate the distance between lithium atoms. For the excited-state geometry, there is a difference between our and \. Zuchowski and Hutson \citep{2010Zuchowski} results. In the case of our computations, the structure is more compact. The Cs-Li distance is significantly smaller (by $\sim$~0.25 \AA), while the Li-Li distance is nearly the same. The atomization energies were calculated by separation of all atoms to the distances at which energy did not change. To confirm that the internuclear distances are large enough, we performed FCI computations, that yielded the same energy as MRCI. The value of the atomization energy (Table \ref{global}) for 1$^2$A$'$ state from our calculations is larger than the one from the most recent study \citep{2010Zuchowski}, but at the same time, it is smaller than the one from the earlier paper \citep{1982Richtsmeier}. In the case of 2$^2$A$'$ state, our atomization energy is again larger in the comparison with value provided by \. Zuchowski and Hutson \citep{2010Zuchowski}, while this energy is not reported by Richtsmeier {\it et al.} \citep{1982Richtsmeier}. These results are rather expected since the D$_e$ value for the LiCs dimer was already larger than in \citep{2010Zuchowski} and we have also used augmented basis sets. On the other hand, the earliest study \citep{1982Richtsmeier} employed the diatomics-in-molecules technique with parameters derived from experiments on the homonuclear diatomic molecules, since at that time no adequate experimental data were available for heteronuclear diatomics. The FCI corrections are small (less than 0.3\%) and suggest, that the choice of active space is reasonable and the MRCI calculations retrieve correlation effects very well. The difference between the ground and first excited doublet A$'$ states energies in the geometric minima, in our and previous results are very close -- the difference in \citep{2010Zuchowski} is 891 cm$^{-1}$ and in our calculations, it is 889 cm$^{-1}$ with FCI correction and 894 cm$^{-1}$ without. Interestingly, it appears that when the Li$_2$Cs trimer is excited from the ground to the first excited state, the equilibrium geometry bond length between the Li and Cs atoms elongates and the one between Li and Li shortens.

\begin{table}[ht]
\caption{Atomization energies and equilibrium parameters of the ground and first excited doublet states of the Li$_2$Cs trimer. Results for the 1$^2$B$_2$ (1$^2$A$'$) are compared with results from \citep{1982Richtsmeier} and \citep{2010Zuchowski}, whereas 1$^2$A$_1$ (2$^2$A$'$) is compared with results from the supplementary materials therein \citep{2010Zuchowski_suppl}. Bond lengths are in \AA~and energies are in cm$^{-1}$ units. This part of calculations were performed with all bond lengths released, but since the equilibrium geometries for both states turned out to be isosceles triangles, only two interatomic distances are given.}
\label{global}
\begin{center}
\begin{tabular}{l c c c c c}
& \multicolumn{3}{c}{Ground state} & \multicolumn{2}{c}{Excited state} \\
& \multicolumn{3}{c}{1$^2$B$_2$ (1$^2$A$'$)} & \multicolumn{2}{c}{1$^2$A$_1$ (2$^2$A$'$)} \\
& present & \citep{2010Zuchowski} & \citep{1982Richtsmeier} & present & \citep{2010Zuchowski_suppl} \\
Bond lengths &  &  &  &  &\\
R (Cs-Li)  & 3.75 & 3.8 & 3.97 & 4.05 & 4.3 \\
R (Li-Li)  & 3.07 & 3.1 & 2.67 & 2.72 & 2.7$^*$ \\
Energies &  &  &  &  \\
Atomization energies  & 11390 & 11073 & 11507 & 10496 & 10182 \\
FCI correction & 25 & --- & --- & 30 & --- \\
\end{tabular}
\footnotesize{\item[*] We derived this value from given data using the law of sines.}
\end{center}
\end{table}

In order to assess the contribution of the nonadditive part of the triatomic interaction energy in the global minimum of the ground state of the Li$_2$Cs molecule, we calculated the interaction energies of the LiCs and Li$_2$ dimers. The interatomic distances used in these calculations R(Cs-Li) and R(Li-Li) (see Table \ref{global}) yielded energies 6037 and 7779 cm$^{-1}$ for LiCs and Li$_2$, respectively. Using equation (\ref{eq:tbnaf}), where $r_{12} = r_{13}$ is the R(Cs-Li) and $r_{23}$ is R(Li-Li), we obtained the value of nonadditive contribution V$_3$ which equals -8462~cm$^{-1}$. It turns out, that since this energy significantly contributes to the atomization energy of the trimer, the simple additive model, that neglects V$_3$, is not suitable for the description of the ground doublet state of the considered trimer.

\begin{figure}[ht]
\begin{center}
\includegraphics[width=14cm]{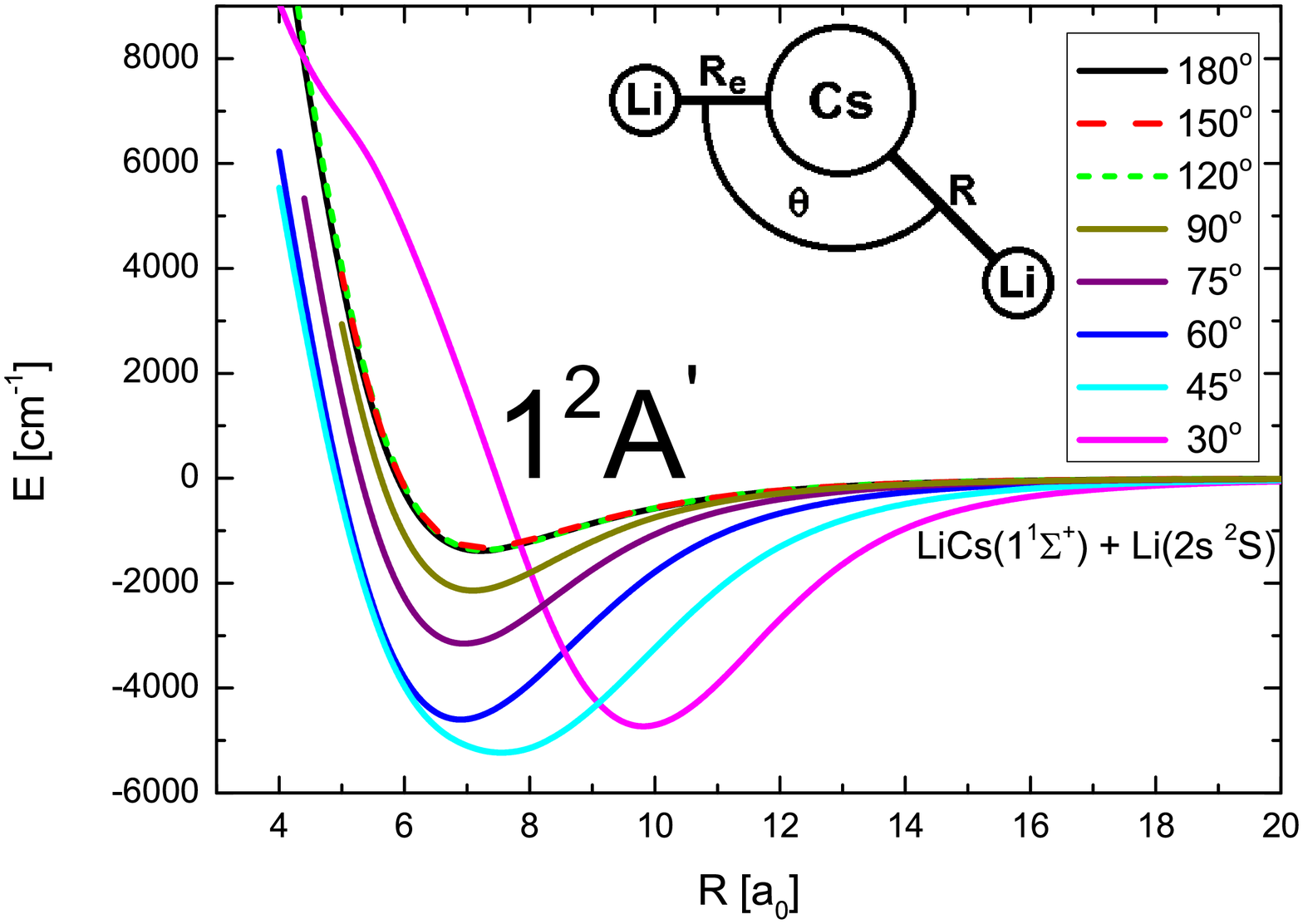}
\caption{\label{LCL} BO adiabatic potential energy curves of the ground state of the system in the LiCs-Li geometry, calculated for eight different angles.}
\end{center}
\end{figure}

\begin{figure}[ht]
\begin{center}
\includegraphics[width=14cm]{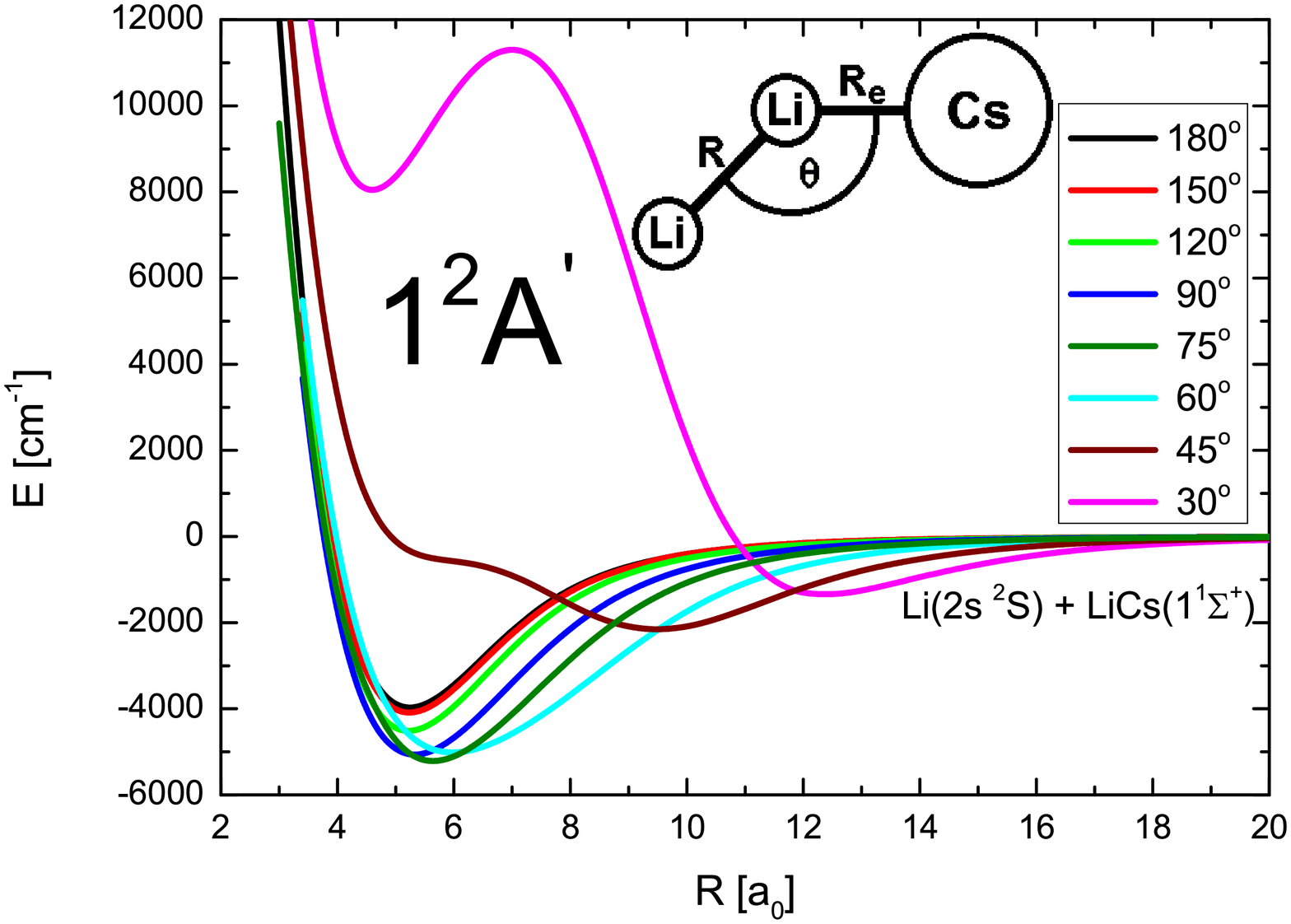}
\caption{\label{CLL} BO adiabatic potential energy curves of the ground state of the system in the Li-LiCs geometry, calculated for eight different angles.}
\end{center}
\end{figure}

\begin{figure}[ht]
\begin{center}
\vspace{-3cm}
\includegraphics[width=14.0cm]{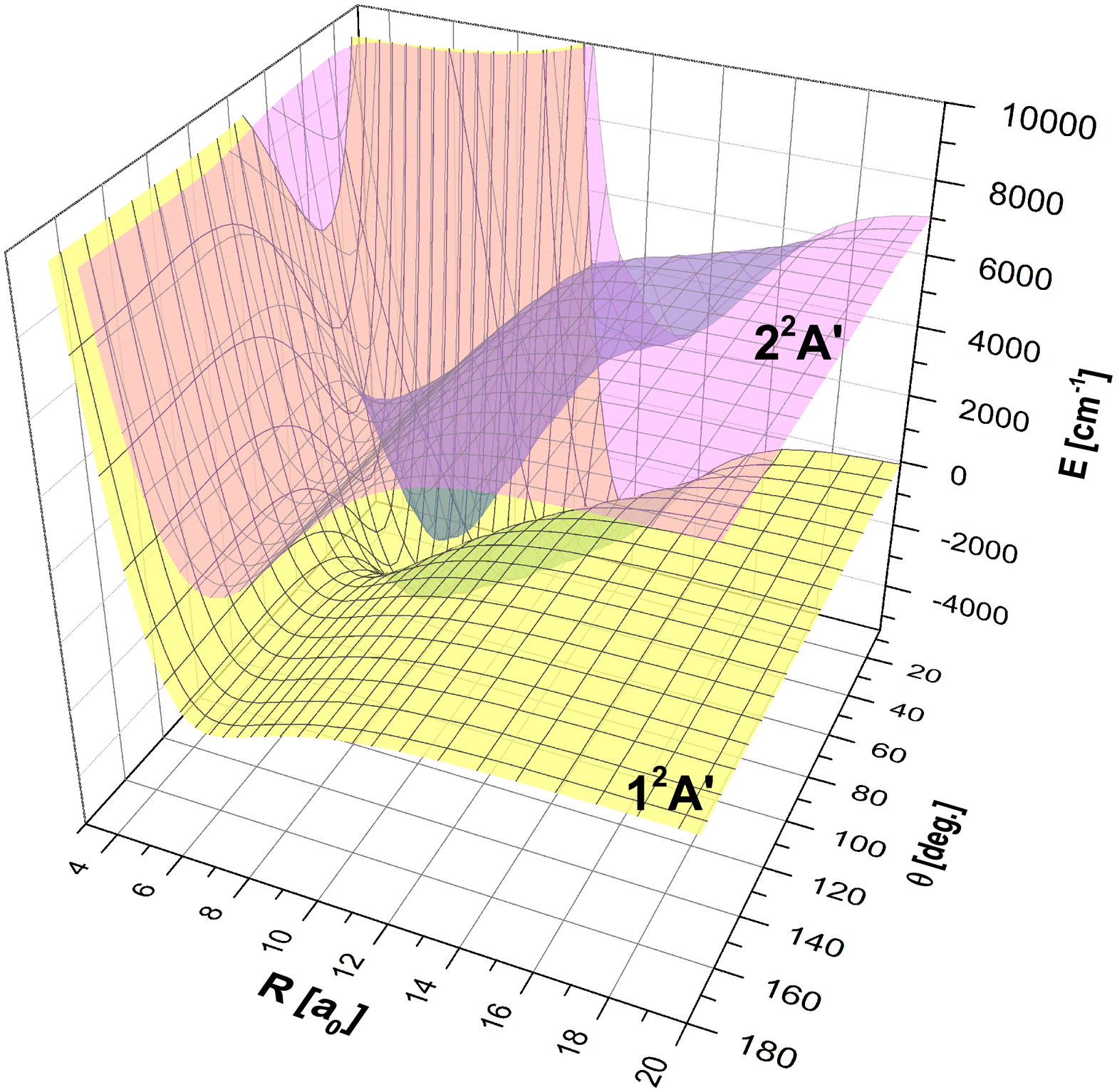} \\
\includegraphics[width=14.0cm]{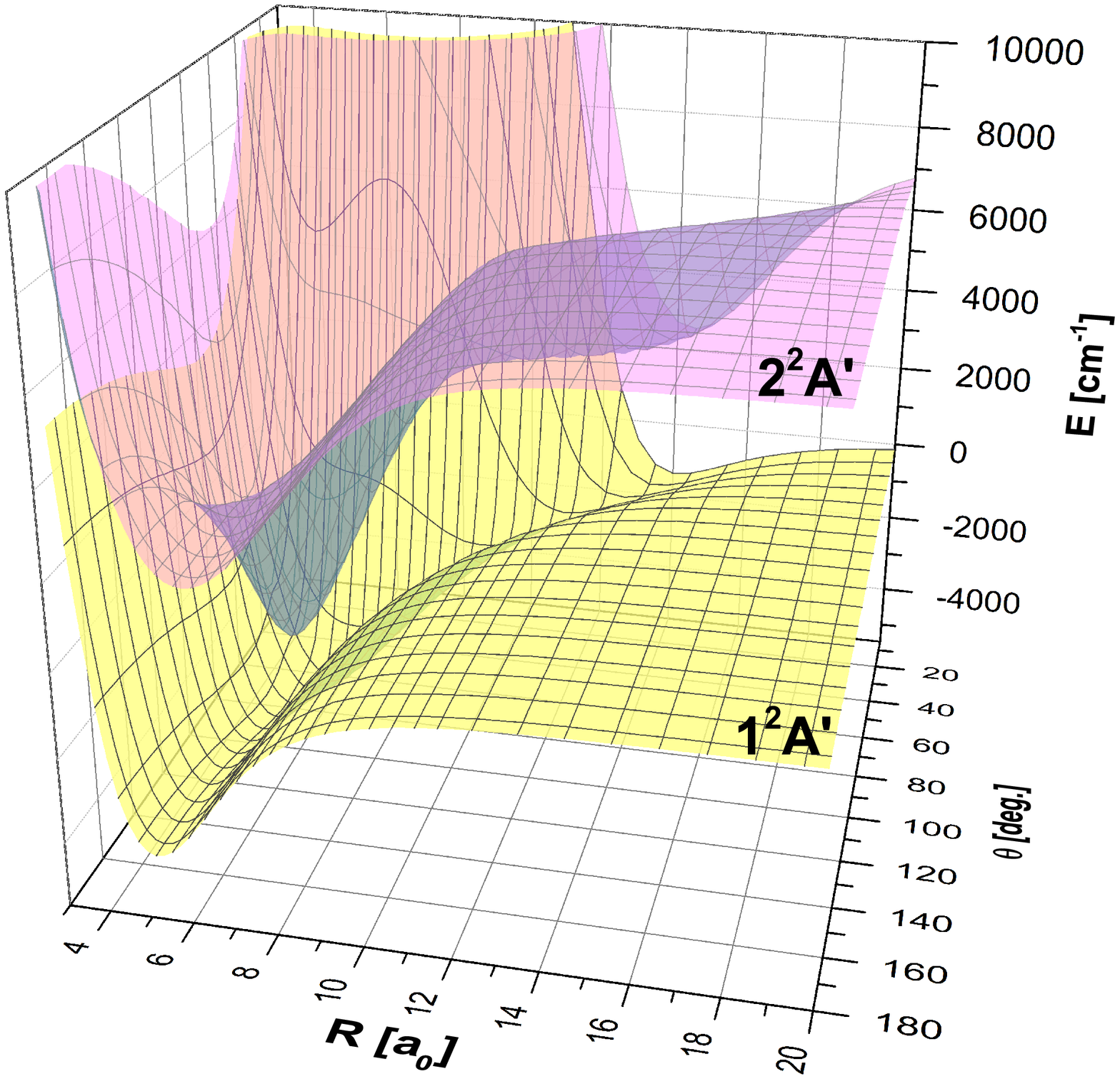}
\caption{\label{LCL-CLL-APES} BO adiabatic potential energy surfaces of the ground 1$^2$A$'$ state and first excited 2$^2$A$'$ state of the LiCs-Li (upper panel) and Li-LiCs (lower panel) geometries of the system.}
\end{center}
\end{figure}

Calculated BO adiabatic potentials for different states of the Li atom interacting with the LiCs molecule are presented in Figs \ref{LCL}-\ref{LCL-CLL-APES}. Particularly, comparing Fig. \ref{LCL} to Fig. \ref{CLL} one may find a difference in the BO adiabatic potentials for the two considered geometries. As one might expect, when the lithium atom approaches the dimer from the side of the lithium atom (Fig. \ref{CLL}), starting from the bond angle $\theta$ = 45$^o$, there is much more pronounced shape of the potential with a characteristic saddle point around R = 7 a$_0$ than in the case shown in Fig. \ref{LCL}. In Fig. \ref{CLL}, for $\theta$ = 30$^o$, we observe the appearance of the two minima, one approximately at R equal to 5 a$_0$ and another close to R = 12 a$_0$. Generally, it is quite obvious that with decrease of the bond angle $\theta$ the potentials' minima shift towards the higher values of R. The reason is that at the short distances the repulsive interaction between the approaching lithium atom and the nearest of dimer's atom grows when $\theta$ becomes smaller.

The BO adiabatic potential energy surfaces of the considered geometries of the Li+LiCs system are shown in Fig \ref{LCL-CLL-APES}, as well as in the figures with the contour plots placed in the supplementary materials \citep{supplementary}. Analysis of these two surfaces allows us to find minima of the ground state 1$^2$A$'$ and the first excited state 2$^2$A$'$ of the lithium atom interacting with the LiCs dimer. The Cs-Li bond length becomes longer after shifting from the minimum of the ground state to the minimum of the excited state, and the Li-Li bond becomes shorter as seen in Table \ref{parameters}. This is analogous to trimer behavior as shown in Table \ref{global}. Additionally, the dissociation energies D$_e$ (Table \ref{parameters}), calculated with respect to the Li(2s~$^2$S)+LiCs(1$^1\Sigma^+$) and Li(2s~$^2$S)+ LiCs(1$^3\Sigma^+$) channels are equal respectively to 5300 and 11066 cm$^{-1}$ for the 1$^2$A$'$ and 2$^2$A$'$ states.

\begin{table}
\caption{\label{parameters} The geometry parameters of Li + LiCs system's minima for the ground and first excited $^2$A$'$ states and appropriate dissociation energies D$_e$. Fixed bond length R$_e$ of the LiCs molecule is equal to 3.6681 \AA. Bond lengths are in \AA~and energies are in cm$^{-1}$ units. The superscript $i$ in $Li^{(i)}$ enumerates lithium atoms.}
\begin{center}
\begin{tabular}{l c c}
 & Ground state & Excited state \\
 & 1$^2$A$'$ & 2$^2$A$'$ \\

Bond lengths &  &  \\
R$_e$ (Cs-Li$^{(2)}$) & 3.6681 & 3.6681 \\
R (Cs-Li$^{(1)}$) & 3.79 & 3.83 \\
R (Li$^{(2)}$-Li$^{(1)}$) & 3.06 & 2.66 \\

Energies &  &  \\
Dissociation energies & 5300 & 11066 \\
FCI correction & 25 & 35 \\
\end{tabular}
\end{center}
\end{table}

\begin{figure}[ht]
\begin{center}
\vspace{-4.5cm}
\includegraphics[width=12.0cm]{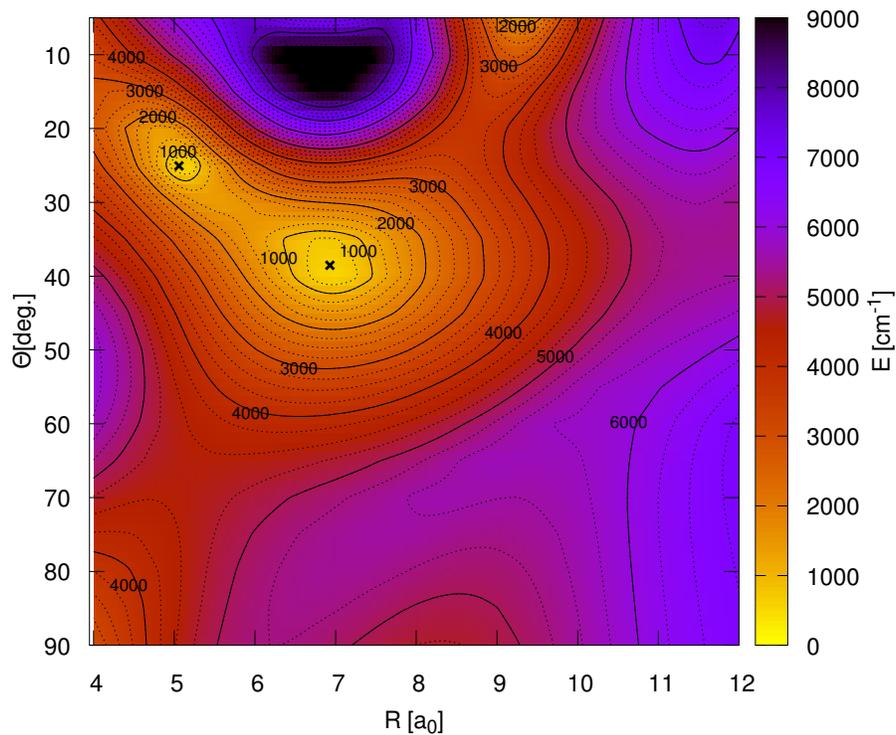} \\
\vspace{-2.0cm}
\includegraphics[width=12.0cm]{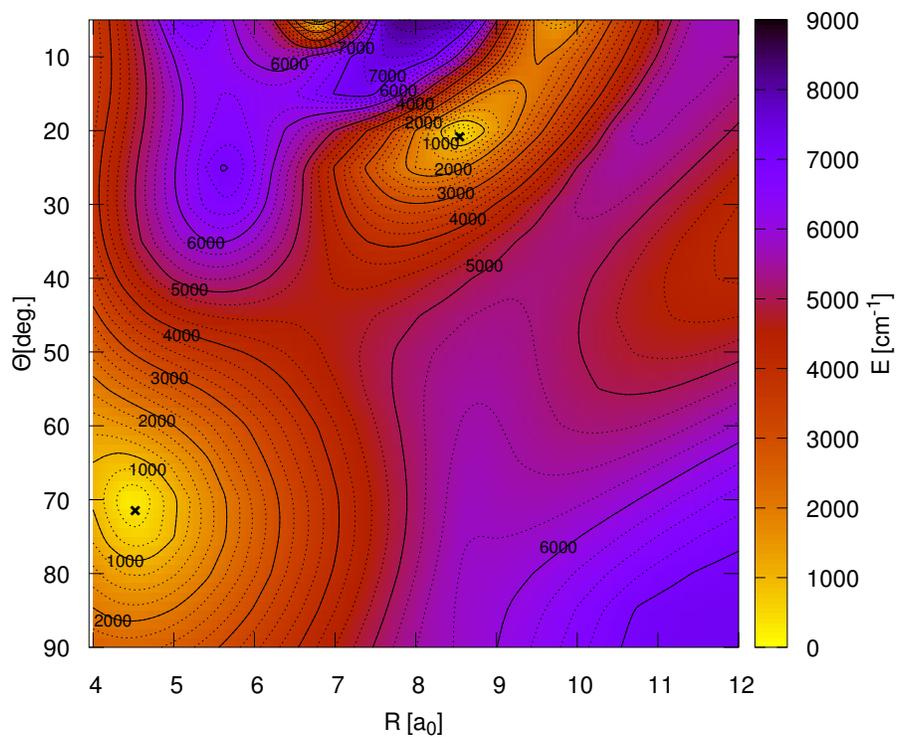}
\vspace{-1.0cm}
\caption{\label{diff-APES} Contour plots of differences (in cm$^{-1}$) between BO adiabatic potential energy surfaces of the ground $1^2A'$ state and first excited $2^2A'$ state of the LiCs-Li (upper panel) and Li-LiCs (lower panel) geometries of the system. The conical intersection points are marked by \textbf{x}.}
\end{center}
\end{figure}

As long as we keep the distance R$_e$ between atoms in the LiCs molecule fixed, we are able to investigate the conical intersections (CI) of the potential energy surfaces in both considered approaches (LiCs-Li and Li-LiCs) as single points only. The three of such CI points are found. Figure \ref{diff-APES} presents colored contour plots of differences between BO adiabatic PESs of the ground 1$^2$A$'$ state and the first excited 2$^2$A$'$ state of the LiCs-Li (upper panel) and Li-LiCs (lower panel) geometries of the system, where conical intersection points are clearly visible. The most important conical intersection is the point lying on the seam in the C$_{2v}$ symmetry. This symmetry introduces the interplay between 1$^2$B$_2$ and 1$^2$A$_1$ electronic states corresponding to 1$^2$A$'$ and 2$^2$A$'$ in the C$_s$ symmetry, i.e. the two lowest lying potential energy surfaces. In our case, we find this conical intersection by setting the distance between the LiCs molecule and the approaching lithium atom to the value equal R$_e$ and finding the angle, for which such calculated potential energy curves cross. The CI determined in that way is visible in both geometries (both panels), as shown in Fig~\ref{diff-APES}. In this point the Li + LiCs system is the isosceles triangle with angles equal to 38.10 (Li-Cs-Li, upper panel) and 70.95 (Li-Li-Cs, lower panel) degrees, as well as bond lengths amount to 6.9317 (LiCs-Li) and 4.5241 (Li-LiCs) a$_0$. The energy of this CI equals 3264 cm$^{-1}$ below the Li(2s~$^2$S) + LiCs(1$^1\Sigma^+$) dissociation limit. The two other conical intersections are situated in the repulsive parts of the PESs. The first one is visible in the upper panel of Fig \ref{diff-APES} where the angle Li-Cs-Li equals 24.13 degrees and the distance LiCs-Li is 5.0640 a$_0$. It is not visible in the lower panel since in the Li-LiCs geometry coordinates, equal to 41.86 degrees and 3.1021 a$_0$, fall out of the shown range. The second one (see the lower panel) is present at the angle Li-Li-Cs equals 20.35 degrees and distance Li-LiCs equals 8.5253 a$_0$. This one is not visible in the upper panel, since its coordinates for the LiCs-Li geometry, equal to 109.70 degrees and 3.1490 a$_0$, are out of the presented range. The respective energies of these conical intersections are 14331 and 18713 cm$^{-1}$ above the Li(2s~$^2$S) + LiCs(1$^1\Sigma^+$) dissociation limit.

\begin{figure}[ht]
\begin{center}
\includegraphics[width=6.55cm]{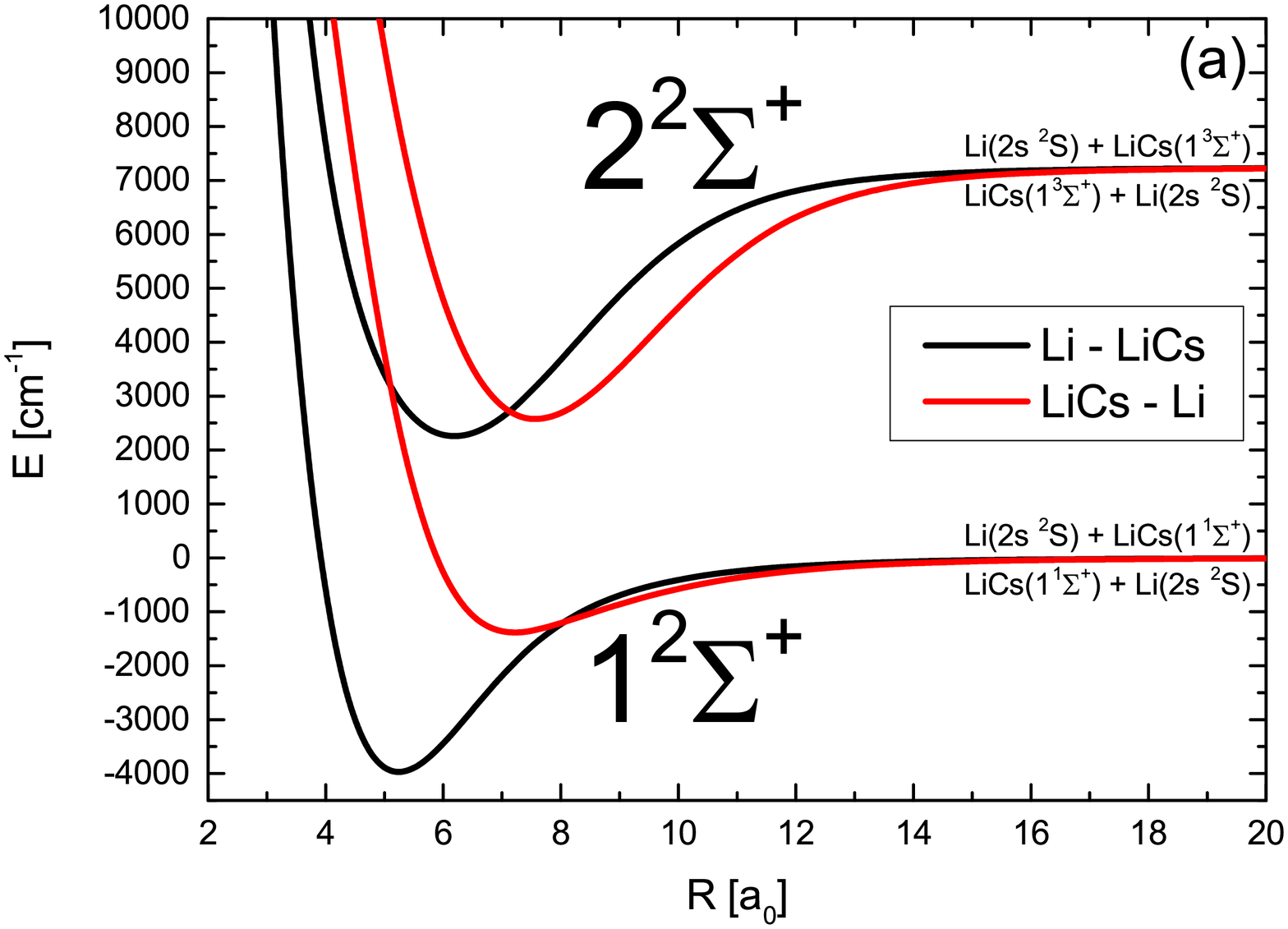} \hspace{-1.2cm} \includegraphics[width=6.55cm]{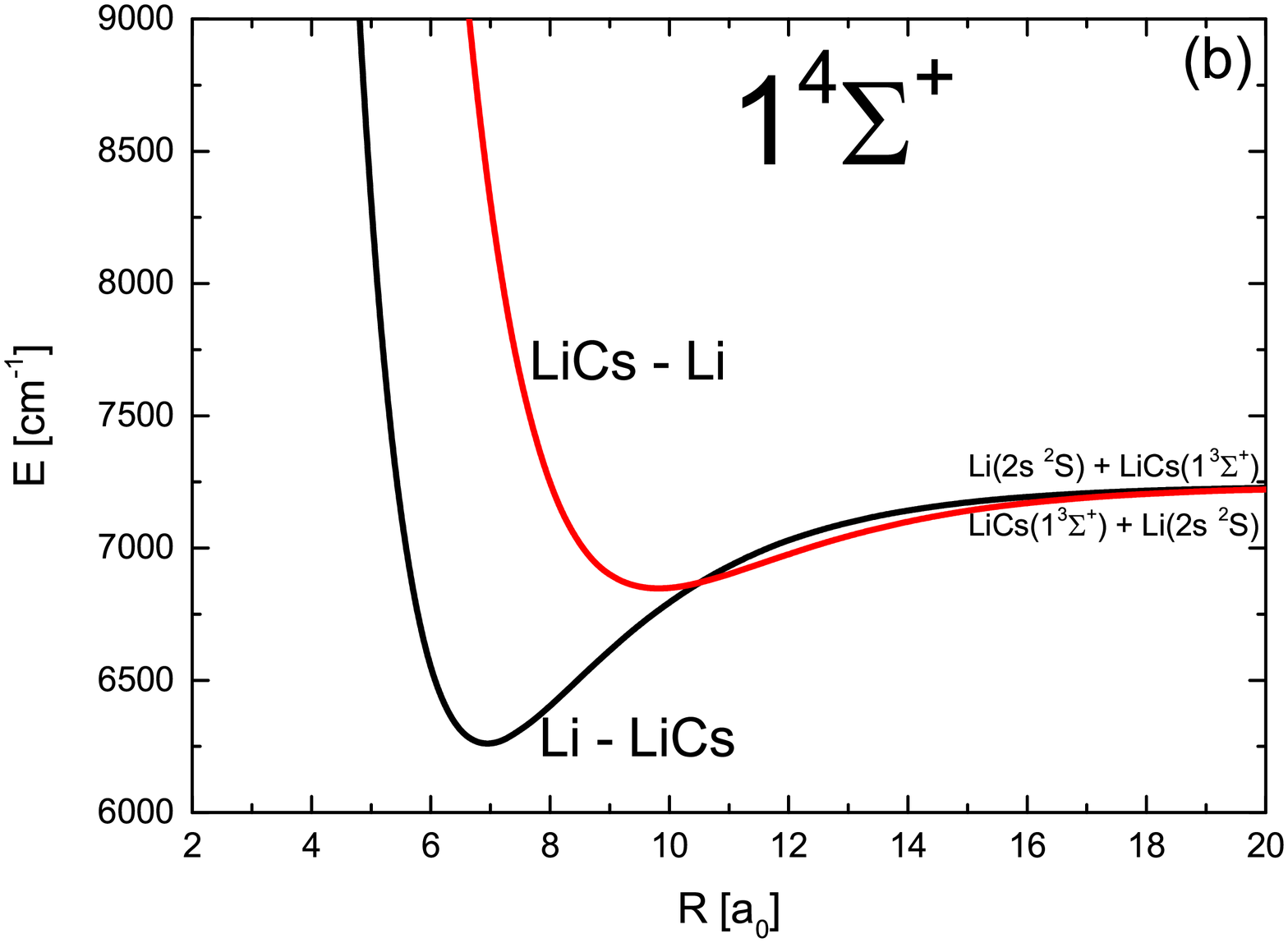} \\
\vspace{-0.5cm}
\includegraphics[width=6.55cm]{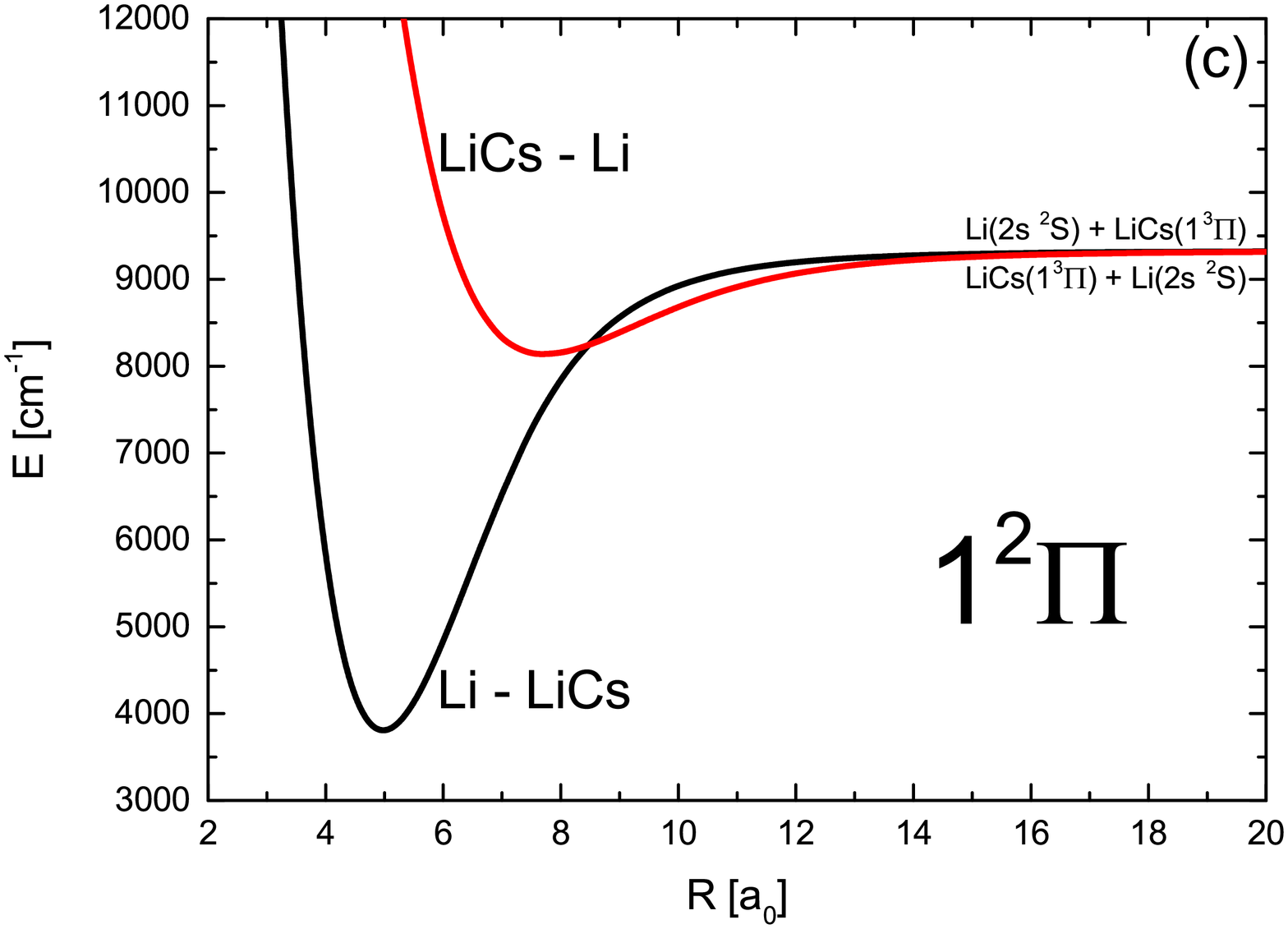} \hspace{-1.2cm} \includegraphics[width=6.55cm]{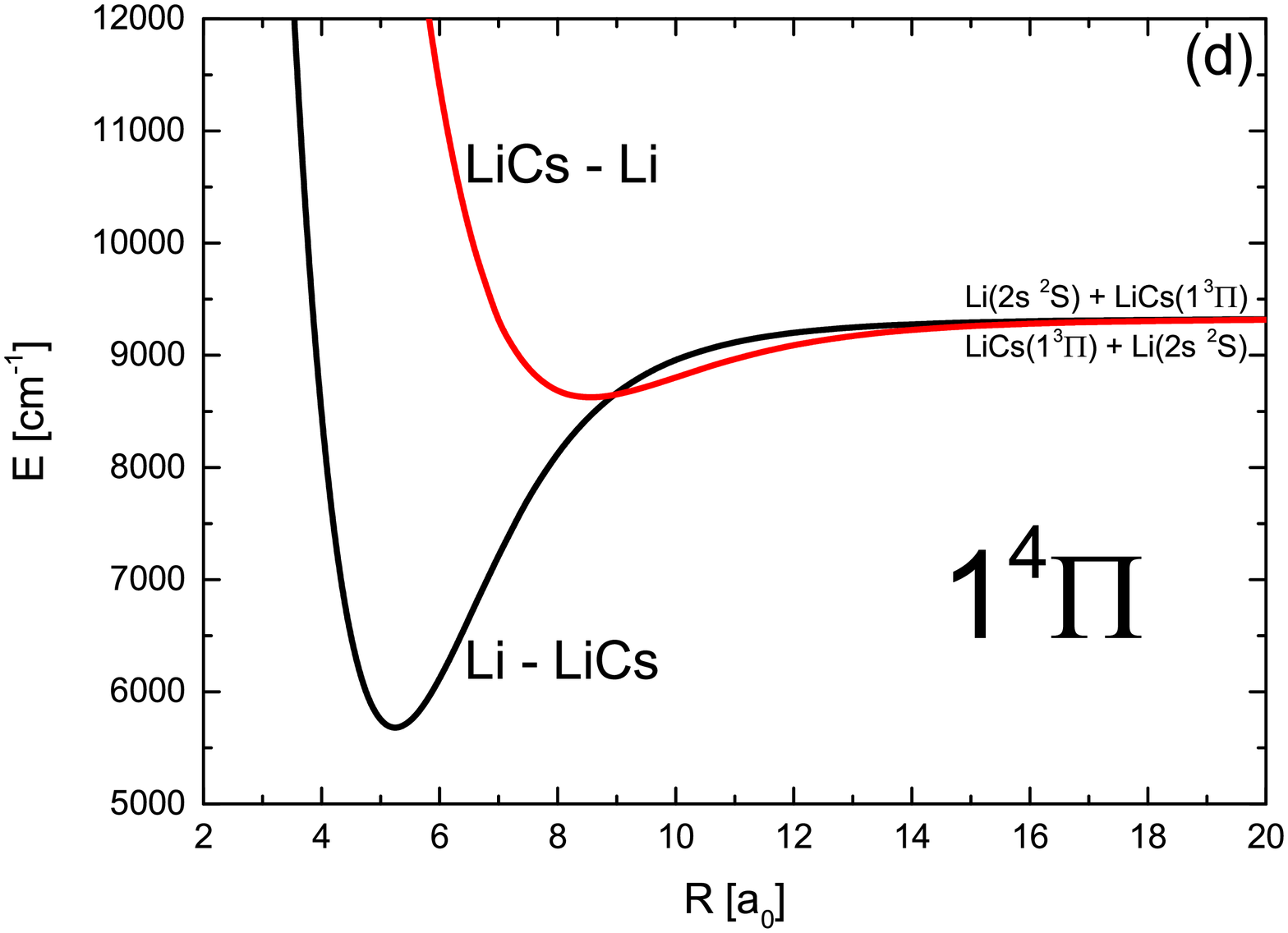} 
\caption{\label{linear} BO adiabatic potential energy curves of the 1,2$^2\Sigma^+$ (a), 1$^4\Sigma^+$ (b), 1$^2\Pi$ (c) and 1$^4\Pi$ (d) states of the two linear geometries. The red and black lines refer to LiCs-Li and Li-LiCs geometries, respectively.}
\end{center}
\end{figure}

Fig \ref{linear} presents the BO adiabatic potentials of the first two doublet $\Sigma^+$ states (a), first quartet $\Sigma^+$  states (b), first doublet $\Pi$ states (c) and first quartet $\Pi$ states (d) of the linear Li + LiCs system for the two considered geometrical cases with the interatomic distance R$_e$ in the LiCs molecule fixed. These potential energy curves correlate with the three lowest-lying dissociation limits. The 1$^2\Sigma^+$ states correlate to the asymptote where both the lithium atom and the LiCs molecule are in the ground states, namely Li(2s~$^2$S) and LiCs(1$^1\Sigma^+$), respectively. The dissociation limit for the 2$^2\Sigma^+$ and 1$^4\Sigma^+$ states corresponds to the sum of the first excited triplet state (1$^3\Sigma^+$) energy of the LiCs molecule and the ground state energy of the lithium atom. Finally, the 1$^2\Pi$ and 1$^4\Pi$ states correlate with the dissociation limit, where the Li atom is in the 2s~$^2$S state, whereas the LiCs molecule is in the 1$^3\Pi$ state.

All potentials presented in Fig \ref{linear} show smooth behavior. It is clearly visible that the differences in the positions of the minima and their depths depend on the atoms arrangement. In the case of LiCs-Li arrangement, the repulsive interaction between the approaching Li atom and the Cs atom of the diatomic molecule is much stronger than in the case of Li-LiCs, where we deal with the repulsion between two lithium atoms. Generally, in the case of LiCs-Li geometry, it leads to more shallow potentials and move their equilibrium positions to larger distances between LiCs molecule and Li atom. The opposite situation exits for Li-LiCs geometry, where potential energy curves are deeper and their minima lying middlingly 2.4 a$_0$ closer in the comparison with previous case. It is also worth to notice, that in the considered Li + LiCs system the doublet states are usually deeper than the quartet states.

All of our calculated potential energy surfaces, potential energy curves in the case of linear arrangement of the Li + LiCs system and contour plots of presented in the manuscript surfaces are available in the supplementary materials for this article \cite{supplementary} and on the group's website~\cite{website}.

\section{Conclusions}

The results of the {\it ab initio} electronic structure calculations for the Li$_2$Cs trimer, as well as the potential energy surfaces of the Li + LiCs system have been presented. The global minima of the ground and first excited doublet A$'$ states were found and compared with other theoretical results. The three-body nonadditive interactions were calculated for the global minimum of the ground doublet A$'$ state, and it turned out, that the simple additive model of dimer interactions is not suitable for the description of this state. The agreement between the present and previous studies is good, although there are some differences in the geometry of the excited state. 

Analysis of the potential energy surfaces has shown characteristic features of the BO adiabatic potentials describing the approach of the lithium atom to the LiCs dimer in its ground and low-lying excited electronic states. We have found substantial differences between potential energy surfaces calculated for different approach directions of the lithium atom. The three conical intersection points have also been found on these surfaces and thoroughly discussed.

Results of our calculations provide insight into the structure of Li$_2$Cs trimer and interaction between the lithium atom with the LiCs diatomic molecule. It could be useful for researchers dealing with spectroscopic measurements of the three-body alkali systems, experiments involving collisions of alkali atoms with heteronuclear alkali diatomic molecules as well as on the cold and ultracold polar alkali molecules.

\subsection*{Acknowledgments}
{We acknowledge partial support by the CMST COST Action CM1204 of the European Community and the Polish Ministry of Science and Higher Education. Calculations have been carried out at the Academic Computer Centre in Gda\' nsk (http://task.gda.pl) and using resources provided by Wroc{\l}aw Centre for Networking and Supercomputing (http://wcss.pl).}

\quad

\bibliography{li+lics_bibliography}

\end{document}